\documentclass[sigconf,nonacm]{acmart}

\usepackage{smartdiagram}
\usepackage{tikz}
\usepackage{pgfplots}
\usetikzlibrary{
    patterns,
}

\usepackage{booktabs}
\usepackage{bm}
\usepackage{caption}
\usepackage{subfigure}
\usepackage{color}
\usepackage{multirow}
\usepackage{amsmath}
\usepackage{adjustbox }
\usepackage{float}

\usepackage{verbatim}

\usepackage{xcolor}
\usepackage{framed}
\theoremstyle{definition}
\definecolor{formalshade}{rgb}{0.8,0.9,0.95}

\usepackage[strict]{changepage}
\definecolor{darkblue}{rgb}{0.0, 0.0, 0.55}
\definecolor{lightgray}{rgb}{0.929, 0.929, 0.929}

\newenvironment{formal}{%
  \MakeFramed{\advance\hsize-\width\FrameRestore}%
  \noindent\hspace{-4.55pt}
  \begin{adjustwidth}{}{7pt}%
  \vspace{2pt}\vspace{2pt}%
}
{%
  \vspace{2pt}\end{adjustwidth}\endMakeFramed%
}

\definecolor{MyPink}{RGB}{255,178,178}
\definecolor{MyBlue}{RGB}{178,178,255}
\usetikzlibrary{er,positioning}
\usetikzlibrary{er,positioning}

\acmPrice{15.00}
\acmISBN{978-1-4503-XXXX-X/18/06}

\begin{document}

\title{GenRec: Large Language Model for Generative Recommendation}

\settopmatter{authorsperrow=4}

\author{Jianchao Ji}
\affiliation{
  \institution{Rutgers University}
  \city{New Brunswick, NJ}
  \country{US}
}
\email{jianchao.ji@rutgers.edu}

\author{Zelong Li}
\affiliation{ 
  \institution{Rutgers University}
  \city{New Brunswick, NJ}
  \country{US}
}
\email{zelong.li@rutgers.edu}

\author{Shuyuan Xu}
\affiliation{ 
  \institution{Rutgers University}
  \city{New Brunswick, NJ}
  \country{US}
}
\email{shuyuan.xu@rutgers.edu}

\author{Wenyue Hua}
\affiliation{ 
  \institution{Rutgers University}
  \city{New Brunswick, NJ}
  \country{US}
}
\email{wenyue.hua@rutgers.edu}

\author{Yingqiang Ge}
\affiliation{ 
  \institution{Rutgers University}
  \city{New Brunswick, NJ}
  \country{US}
}
\email{yingqiang.ge@rutgers.edu}

\author{Juntao Tan}
\affiliation{ 
  \institution{Rutgers University}
  \city{New Brunswick, NJ}
  \country{US}
}
\email{juntao.tan@rutgers.edu}

\author{Yongfeng Zhang}
\affiliation{
  \institution{Rutgers University}
  \city{New Brunswick, NJ}
  \country{US}
}
\email{yongfeng.zhang@rutgers.edu}

\renewcommand{\shortauthors}{Jianchao Ji et al.}

\begin{abstract}
In recent years, large language models (LLM) have emerged as powerful tools for diverse natural language processing tasks. However, their potential for recommender systems under the generative recommendation paradigm remains relatively unexplored. This paper presents an innovative approach to recommendation systems using large language models (LLMs) based on text data. In this paper, we present a novel LLM for generative recommendation (GenRec) that utilized the expressive power of LLM to directly generate the target item to recommend, rather than calculating ranking score for each candidate item one by one as in traditional discriminative recommendation.
GenRec uses LLM's understanding ability to interpret context, learn user preferences, and generate relevant recommendation. Our proposed approach leverages the vast knowledge encoded in large language models to accomplish recommendation tasks. We first we formulate specialized prompts to enhance the ability of LLM to comprehend recommendation tasks. Subsequently, we use these prompts to fine-tune the LLaMA backbone LLM on a dataset of user-item interactions, represented by textual data, to capture user preferences and item characteristics. Our research underscores the potential of LLM-based generative recommendation in revolutionizing the domain of recommendation systems and offers a foundational framework for future explorations in this field. We conduct extensive experiments on benchmark datasets, and the experiments shows that our GenRec has significant better results on large dataset. Code and data are open-sourced at \url{https://github.com/rutgerswiselab/GenRec}.
\end{abstract}


\maketitle

\section{Introduction}
Large Language Models (LLMs) have made a particularly significant milestone in this technological evolution. These LLMs, designed to understand and generate human-like text, have revolutionized numerous applications, from search engines to chatbots, and have facilitated more natural and intuitive interactions between humans and machines. This paper seeks to explore a relatively new and promising application of these models in the recommendation systems.

Recommendation systems have become an integral part of our digital experience. They are the unseen force guiding us through vast amounts of data, suggesting relevant products on e-commerce websites, recommending movies on streaming platforms, and even proposing what news to read or videos to watch. The primary aim of these systems is to predict the individual user preferences and enhance user experience and engagement.

Traditionally, recommendation systems have been built around methods such as collaborative filtering \cite{he2017neural,konstan1997grouplens,schafer2007collaborative}, content-based filtering \cite{van2000using,son2017content}, and hybrid approaches \cite{basilico2004unifying,pazzani1999framework}. Collaborative filtering leverages user-item interactions, making suggestions based on patterns found in the behavior of similar users or items. On the other hand, content-based filtering uses item features to recommend similar items to those a user has previously interacted with. Hybrid methods attempt to combine the strengths of these two approaches to overcome their respective limitations.

Despite the progress made with these traditional techniques, there still have some significant challenges. For instance, collaborative filtering struggles with the cold start problem, where it fails to provide accurate recommendations for new users or items due to lack of historical interaction data. Both content-based filtering hard to handle the issue of data sparsity, given that most users interact with only a small fraction of the total items available. Additionally, because of  the computational complexity of processing large interaction matrices, these models often struggle to scale effectively with the growth of users and items.

The integration of text-based LLMs into recommendation systems presents an exciting opportunity to address these challenges \cite{geng2022recommendation}. These models can learn and understand complex patterns in human language, which allows for a more nuanced interpretation of user preferences and a more sophisticated generation of recommendations. However,  a significant number of the prevailing recommendation models are trained using user and item indexes. This approach leads to the lack of text-based information in the dataset, including details like item titles and category information.

In this paper, we propose a novel large language model for generative recommendation (GenRec). 
One of the primary benefits of the GenRec model is that it capitalizes on the rich, descriptive information inherently contained within the item names,
which often contain features that can be semantically analyzed, enabling a better understanding of the item's potential relevance to the user. This could potentially provide more accurate and personalized recommendations, thereby enhancing the overall user experience.

We present experimental results to demonstrate the efficacy of our proposed method and compare its performance with other LLM recommendation models. The overarching aim of this paper is not only to present our findings but also to inspire further research in this area. By highlighting the potential of LLMs in enhancing generative recommender systems, we hope to encourage a more widespread adoption of these models and stimulate further innovations in this field.

The key contributions of this paper can be summarized as follows:
\begin{itemize}
\item We highlight the promising paradigm of generative recommendation, which directly generates the target item to recommend, rather than traditional discriminative recommendation, which has to calculate a ranking score for each candidate item one by one and then sorts them for deciding which to recommend.
\item We introduce a novel approach, GenRec, to enhance the generative recommendation performance by incorporating the textual information into the model.
\item We also illustrate the efficacy of GenRec on practical recommendation tasks, underscoring its prospective abilities for a wider scope of applications.
\end{itemize}

In the following parts of the paper, we will discuss the related work in Section \ref{sec:related}, introduce the proposed model in Section \ref{sec:model}, analyze the experimental results in Section \ref{sec:experiment}, and provide the conclusions as well as future work in Section \ref{sec:conclusion}.

\section{Related Work}
\label{sec:related}

\subsection{Collaborative Filtering and Content-Based Recommendation Systems}

Collaborative Filtering (CF) models are based on the concept of user-item interactions. Traditional CF models, such as the matrix factorization model \cite{mnih2007probabilistic}, focus on latent factor modeling of user-item interaction matrices. More recent advancements, like NeuMF \cite{he2017neural}, have combined the merits of matrix factorization and neural networks to better capture complex user-item relationships.

On the other hand, Content-Based Recommendation systems rely on the features of items to make recommendations. Early works involved simple keyword matching \cite{bhalotia2002keyword} or cosine similarity based on TF-IDF vectors \cite{ramos2003using}. More advanced methods have started to exploit deep learning techniques, like CNN \cite{o2015introduction} and RNN \cite{sherstinsky2020fundamentals}, for extracting high-level features from item content.

\subsection{\mbox{Large Language Models for Recommendation}}

The use of large language models for recommendation systems has gained significant attention recently. These models exhibit great potential in the understanding and modeling of user-item interactions, exploiting rich semantics and long-range dependencies present in user activity data.


The pioneering work of P5 \cite{geng2022recommendation} illustrated the feasibility of formulating recommendation as a natural language task. 
P5 \cite{geng2022recommendation} 
fines the widely-used open-source T5 model \cite{raffel2020exploring} to create a unified system capable of handling various tasks. These tasks include not only recommendation ranking and retrieval but also complex functions like summary explanation. This innovative approach highlighted the versatility of large language models in handling multi-task learning in the recommendation context. However, the potential of large language models to understand and generate text-based recommendations has not been fully explored.

In this paper, we propose a novel approach to text-based generative recommendation, leveraging the latest advances in large language models. We aim to address some of the limitations of previous works and push the boundaries of what is possible in the realm of recommendation systems.

\section{Method}
\label{sec:model}
\begin{figure}[t]
    \centering
    \includegraphics[scale=0.43]{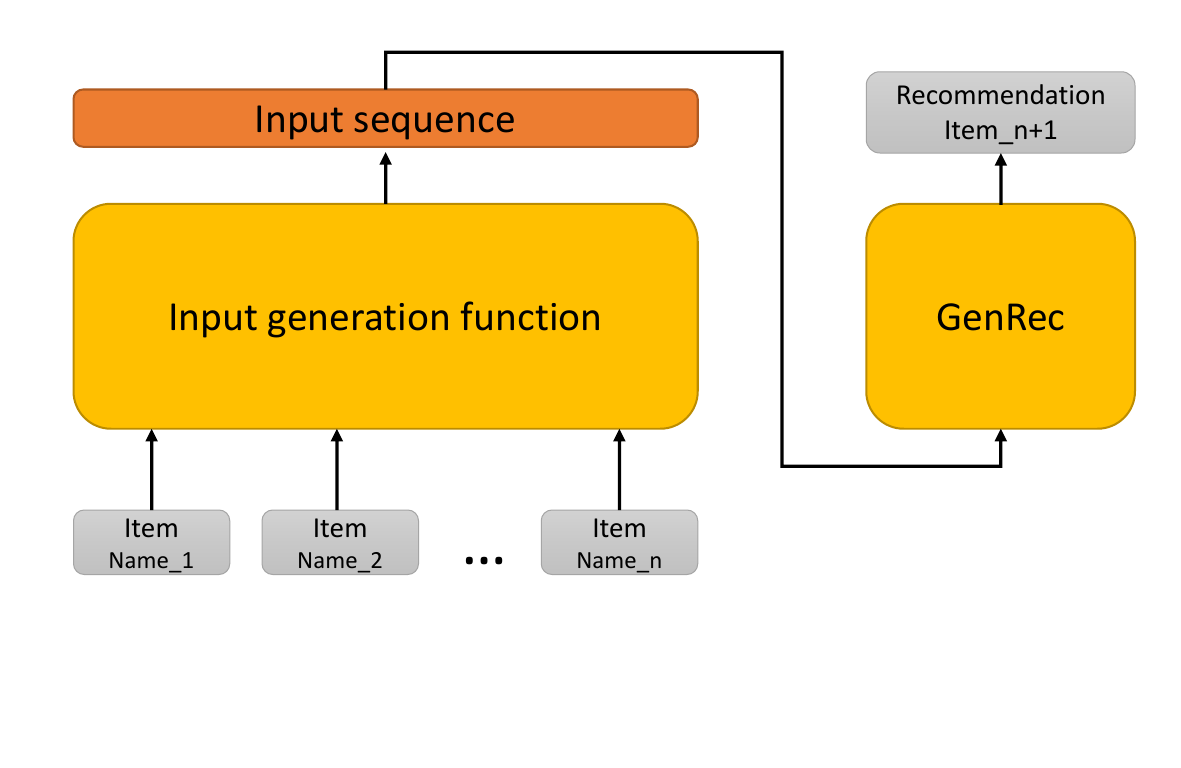}
    \vspace{-50pt}
    \caption{An illustration of GenRec. Our model will generate a input sequence based on the interaction history. Then the model will predict the next item the user may interact with.}
    \vspace{-20pt}
    \label{figure:model}
\end{figure}

The architecture of the proposed framework is illustrated in Figure \ref{figure:model}. Given a user's item interaction sequence, the large language model for generative recommendation (GenRec) will format the item names with a prompt. This reformatted sequence is subsequently employed to fine-tune a Large Language Model (LLM). The adjusted LLM can then predict subsequent items the user is likely to interact with. In our paper, we  select the LLaMA \cite{touvron2023llama} language model as the backbone. However, our framework retains flexibility, allowing for seamless integration with any other LLM, thus broadening its potential usability and adaptability.

\subsection{Sequence Generation}
The initial component of GenRec is a generative function, tasked with producing various sequences that encapsulate user interests.  To enhance the model's comprehension of the recommendation task, we have devised multiple prompts that facilitate sequence generation. Take Figure \ref{figure:recommendation} as an example, we use the user's movie watching history as the training data and use this information to format the training sequence. The sequence consist of three part, instruction input and output. The instruction element outlines the specific task of movie recommendation, for which we have created several directives to enhance the LLM's comprehension of the ongoing recommendation task. The input represents the history of the user's interactions, excluding the most recent instance. And the output is the latest interaction in this record. The primary task for the LLM here is to predict this final interaction accurately.

\begin{table*}[t]
\centering
\begin{adjustbox}{width=0.98\linewidth}
\begin{tabular}{ccccccccc}
\toprule
\multirow{2.5}{*}{Methods} & \multicolumn{4}{c}{\textbf{MovieLens 25M}} & \multicolumn{4}{c}{\textbf{Amazon Toys}}  \\
\cmidrule(lr){2-5}\cmidrule(lr){6-9}
 & HR@5  & NDCG@5 & HR@10  & NDCG@10 & HR@5  & NDCG@5 & HR@10  & NDCG@10 \\
\cmidrule{1-9}
P5   & 0.0688  & 0.0464  & 0.1040 & 0.0577 & $\bm{0.0239}$ & $\bm{0.0145}$ & $\bm{0.0411}$ & $\bm{0.0201}$\\
GenRec    &  $\bm{0.1034}$  & $\bm{0.0716}$  & $\bm{0.1311}$  &  $\bm{0.0837}$ & 0.0190  & 0.0136  & 0.0251  & 0.0157 \\ 
\bottomrule
\end{tabular}
\end{adjustbox}
\caption{Experimental results on Normalize Discounted Cumulative Gain (NDCG@k) and Hit Ratio (HR@k). Bold numbers represent best performance. }
\label{tab:sequential}
\vspace{-20pt}
\end{table*}

\begin{figure}[h]
\begin{formal}
    \textbf{Interaction history}: Pinocchio (1940), Legends of the Fall (1994), Once Were Warriors (1994), In the Name of the Father (1993), Shadowlands (1993), Heavenly Creatures (1994), Quiz Show (1994), In the Line of Fire (1993)\\
    \textbf{Recommendation Prompt Example}:\\
    \textbf{\textit{Instruction}}: Given the movie viewing habits, what is the most probable movie they will choose to watch next?\\
    \textbf{\textit{input}}: Pinocchio (1940), Legends of the Fall (1994), Once Were Warriors (1994), In the Name of the Father (1993), Shadowlands (1993), Heavenly Creatures (1994), Quiz Show (1994)\\
    \textbf{\textit{output}}: In the Line of Fire (1993)\\
\end{formal}
\caption{GenRec on recommendation task. Based on the interactive history, GenRec can convert them to a training sequence which consists of instruction, input and output.}
\label{figure:recommendation}
\vspace{-10pt}
\end{figure}

Refer to Figure \ref{figure:recommendation} for an illustration. This figure represents how we utilize a user's history of watched movies as interaction data. Given the prompt, "Based on the movie viewing habits, what is the most likely movie they will select to watch next?" and the provided input, we then allow GenRec to forecast the subsequent output.

\subsection{Training Strategy}
In this paper, we use the LLaMA model as the backbone for the training of GenRec. The LLaMA model is pre-trained on an expansive language corpus, offering a valuable resource for our intended purpose of efficiently capturing both user interests and item content information. However, it's important to note that the memory requirements for GPU to fine-tune LLaMA, even the 7-billion parameter version, are pretty substantial.

To circumvent this challenge and conserve GPU memory, we adopt the LLaMA-LoRA architecture for fine-tuning and inference tasks within the scope of this study. By this measure, we have achieved a significant reduction in the GPU memory requirements. With this optimized approach, we can fine-tune the LLaMA-LoRA model on a single GPU with a memory capacity of 24GB.

However, in an effort to decrease the overall training time, we have employed a data parallel technique and leveraged multiple GPUs in the experiments. Further details about our experiments, including the implementation and results, will be shared in the following sections of this paper.

\section{Experiments}
\label{sec:experiment}
\subsection{Dataset}
We conduct extensive experiments on two real-world datasets from Amazon \cite{ni2019justifying} and MovieLens \cite{harper2015movielens}, respectively, to evaluate the performance of our proposed GenRec approach on recommendation tasks. The Amazon datasets, which record user purchase histories across a diverse range of products, were sourced from the Amazon.com platform. MovieLens datasets comprise a large number of movie ratings and associated metadata, contributed by users of the MovieLens website over various periods. The descriptive statistics of these datasets are depicted in Table 1 (see reference \ref{tab:stats}). For each user interaction sequence, the most recent item is used as the test data, the second-most recent is used as validation data, and the remaining is used for training.

\begin{table}[h]
\centering
\begin{tabular}{l|r|r}
\toprule
Dataset &  {\textbf{MovieLens 25M}} &  {\textbf{Amazon Toys}}  \\
\cmidrule{1-3}
\#Users    &  162,541  &  19,412   \\
\#Items    &  62,423   &  11,924  \\
\#Interaction &  25,000,095  & 2,252,771  \\
\bottomrule
\end{tabular}
\caption{Basic statistics of the recommendation datasets.}
\vspace{-20pt}
\label{tab:stats}
\end{table}

\subsection{Evaluation Merics}
In this paper, we evaluate the performance of the model using two widely used metrics : Hit Ratio (HR) and Normalized Discounted Cumulative Gain (NDCG). The HR metric indicates the percentage of items recommended by the model that match those in the ground truth data. On the other hand, NDCG is employed to assess the efficacy of the recommendations when they are ranked, factoring in the relevance of the suggested items. These metrics have found wide acceptance in the evaluation of recommendation systems due to their robustness and comprehensiveness.

\subsection{Implement Details}

The GenRec model was pretrained for 5 epochs using the AdamW optimization \cite{loshchilov2018decoupled} on four NVIDIA RTX A5000 GPUs with a batch size of 128. The peak learning rate was set to $3 \times 10^{-4}$ and the maximum input length was set to 256 tokens. A warm-up strategy was employed during training, where the learning rate was gradually increased over the first 1000 steps. 

\subsection{Baseline Methods}

\textbf{P5} \cite{geng2022recommendation}: The Pre-train, Personalized Prompt, and Predict Paradigm (P5) incorporates an array of templates for input and target sequences throughout the training process. This unique approach proficiently dissolves the boundaries between different tasks, promoting a more fluid and integrated training procedure. It has showcased noteworthy performance in the domain of sequential recommendation tasks, underlining its effectiveness and applicability.

\subsection{Performance Comparison}

As we can see in the Table \ref{tab:sequential}, P5 has better performance on Amazon Toys datasets, while our GenRec has significant better performance on movielens 25M datasets. The possible reasons behind this differential performance could be attributed to the distinct nature of the datasets. The MovieLens 25M dataset, unlike Amazon Toys datasets, contains a richer amount of interaction information, which provides a more robust understanding of the user's preferences and behavior, thus likely leading to more accurate recommendations.

Our GenRec model, designed to effectively capture both user interests and item content information and produce more accurate and relevant recommendations. On the other hand, P5, while robust in handling sequential data, might not be as adept in leveraging this additional interaction information, resulting in relatively lower performance on the MovieLens 25M dataset.

\section{Conclusion}
In conclusion, our work on the text-based Large Language Model for Generative Recommendation (GenRec) has revealed a novel and promising approach in the field of recommendation systems. By focusing on the semantic richness of item names as input, GenRec promises more personalized and contextually relevant recommendations. Our practical demonstrations highlight GenRec's efficacy and point towards its adaptability across a diverse range of applications. Furthermore, the flexibility of the GenRec framework facilitates integration with any Large Language Model, hence widening its sphere of potential utility.

In terms of future work, there are several directions to explore. We intend to refine the generation of sequences by developing more sophisticated prompts, which could further enhance the model's understanding of recommendation tasks. Additionally, we plan to extend our research to incorporate more complex user interaction data, such as ratings or reviews, which could provide deeper insights into user behavior and preferences. A further direction would be to test GenRec's performance with different Large Language Models, investigating the possible benefits and trade-offs.

Our research with GenRec thus far has shown significant promise, and we look forward to continuing to develop and refine this approach. We believe that with further investigation, GenRec could revolutionize the way recommendation systems operate, ultimately leading to more personalized and satisfying user experiences.

\label{sec:conclusion}
\bibliographystyle{ACM-Reference-Format}
\balance
\bibliography{reference.bib}


\begin{thebibliography}{18}


\ifx \showCODEN    \undefined \def \showCODEN     #1{\unskip}     \fi
\ifx \showDOI      \undefined \def \showDOI       #1{#1}\fi
\ifx \showISBNx    \undefined \def \showISBNx     #1{\unskip}     \fi
\ifx \showISBNxiii \undefined \def \showISBNxiii  #1{\unskip}     \fi
\ifx \showISSN     \undefined \def \showISSN      #1{\unskip}     \fi
\ifx \showLCCN     \undefined \def \showLCCN      #1{\unskip}     \fi
\ifx \shownote     \undefined \def \shownote      #1{#1}          \fi
\ifx \showarticletitle \undefined \def \showarticletitle #1{#1}   \fi
\ifx \showURL      \undefined \def \showURL       {\relax}        \fi
\providecommand\bibfield[2]{#2}
\providecommand\bibinfo[2]{#2}
\providecommand\natexlab[1]{#1}
\providecommand\showeprint[2][]{arXiv:#2}

\bibitem[Basilico and Hofmann(2004)]%
        {basilico2004unifying}
\bibfield{author}{\bibinfo{person}{Justin Basilico} {and}
  \bibinfo{person}{Thomas Hofmann}.} \bibinfo{year}{2004}\natexlab{}.
\newblock \showarticletitle{Unifying collaborative and content-based
  filtering}. In \bibinfo{booktitle}{\emph{Proceedings of the twenty-first
  international conference on Machine learning}}. \bibinfo{pages}{9}.
\newblock


\bibitem[Bhalotia et~al\mbox{.}(2002)]%
        {bhalotia2002keyword}
\bibfield{author}{\bibinfo{person}{Gaurav Bhalotia}, \bibinfo{person}{Arvind
  Hulgeri}, \bibinfo{person}{Charuta Nakhe}, \bibinfo{person}{Soumen
  Chakrabarti}, {and} \bibinfo{person}{Shashank Sudarshan}.}
  \bibinfo{year}{2002}\natexlab{}.
\newblock \showarticletitle{Keyword searching and browsing in databases using
  BANKS}. In \bibinfo{booktitle}{\emph{Proceedings 18th international
  conference on data engineering}}. IEEE, \bibinfo{pages}{431--440}.
\newblock


\bibitem[Geng et~al\mbox{.}(2022)]%
        {geng2022recommendation}
\bibfield{author}{\bibinfo{person}{Shijie Geng}, \bibinfo{person}{Shuchang
  Liu}, \bibinfo{person}{Zuohui Fu}, \bibinfo{person}{Yingqiang Ge}, {and}
  \bibinfo{person}{Yongfeng Zhang}.} \bibinfo{year}{2022}\natexlab{}.
\newblock \showarticletitle{Recommendation as language processing (rlp): A
  unified pretrain, personalized prompt \& predict paradigm (p5)}. In
  \bibinfo{booktitle}{\emph{Proceedings of the 16th ACM Conference on
  Recommender Systems}}. \bibinfo{pages}{299--315}.
\newblock


\bibitem[Harper and Konstan(2015)]%
        {harper2015movielens}
\bibfield{author}{\bibinfo{person}{F~Maxwell Harper} {and}
  \bibinfo{person}{Joseph~A Konstan}.} \bibinfo{year}{2015}\natexlab{}.
\newblock \showarticletitle{The movielens datasets: History and context}.
\newblock \bibinfo{journal}{\emph{Acm transactions on interactive intelligent
  systems (tiis)}} \bibinfo{volume}{5}, \bibinfo{number}{4}
  (\bibinfo{year}{2015}), \bibinfo{pages}{1--19}.
\newblock


\bibitem[He et~al\mbox{.}(2017)]%
        {he2017neural}
\bibfield{author}{\bibinfo{person}{Xiangnan He}, \bibinfo{person}{Lizi Liao},
  \bibinfo{person}{Hanwang Zhang}, \bibinfo{person}{Liqiang Nie},
  \bibinfo{person}{Xia Hu}, {and} \bibinfo{person}{Tat-Seng Chua}.}
  \bibinfo{year}{2017}\natexlab{}.
\newblock \showarticletitle{Neural collaborative filtering}. In
  \bibinfo{booktitle}{\emph{Proceedings of the 26th international conference on
  world wide web}}. \bibinfo{pages}{173--182}.
\newblock


\bibitem[Konstan et~al\mbox{.}(1997)]%
        {konstan1997grouplens}
\bibfield{author}{\bibinfo{person}{Joseph~A Konstan},
  \bibinfo{person}{Bradley~N Miller}, \bibinfo{person}{David Maltz},
  \bibinfo{person}{Jonathan~L Herlocker}, \bibinfo{person}{Lee~R Gordon}, {and}
  \bibinfo{person}{John Riedl}.} \bibinfo{year}{1997}\natexlab{}.
\newblock \showarticletitle{Grouplens: Applying collaborative filtering to
  usenet news}.
\newblock \bibinfo{journal}{\emph{Commun. ACM}} \bibinfo{volume}{40},
  \bibinfo{number}{3} (\bibinfo{year}{1997}), \bibinfo{pages}{77--87}.
\newblock


\bibitem[Loshchilov and Hutter(2019)]%
        {loshchilov2018decoupled}
\bibfield{author}{\bibinfo{person}{Ilya Loshchilov} {and}
  \bibinfo{person}{Frank Hutter}.} \bibinfo{year}{2019}\natexlab{}.
\newblock \showarticletitle{Decoupled Weight Decay Regularization}. In
  \bibinfo{booktitle}{\emph{International Conference on Learning
  Representations}}.
\newblock
\urldef\tempurl%
\url{https://openreview.net/forum?id=Bkg6RiCqY7}
\showURL{%
\tempurl}


\bibitem[Mnih and Salakhutdinov(2007)]%
        {mnih2007probabilistic}
\bibfield{author}{\bibinfo{person}{Andriy Mnih} {and} \bibinfo{person}{Russ~R
  Salakhutdinov}.} \bibinfo{year}{2007}\natexlab{}.
\newblock \showarticletitle{Probabilistic matrix factorization}.
\newblock \bibinfo{journal}{\emph{Advances in neural information processing
  systems}}  \bibinfo{volume}{20} (\bibinfo{year}{2007}).
\newblock


\bibitem[Ni et~al\mbox{.}(2019)]%
        {ni2019justifying}
\bibfield{author}{\bibinfo{person}{Jianmo Ni}, \bibinfo{person}{Jiacheng Li},
  {and} \bibinfo{person}{Julian McAuley}.} \bibinfo{year}{2019}\natexlab{}.
\newblock \showarticletitle{Justifying recommendations using distantly-labeled
  reviews and fine-grained aspects}. In \bibinfo{booktitle}{\emph{Proceedings
  of the 2019 conference on empirical methods in natural language processing
  and the 9th international joint conference on natural language processing
  (EMNLP-IJCNLP)}}. \bibinfo{pages}{188--197}.
\newblock


\bibitem[O'Shea and Nash(2015)]%
        {o2015introduction}
\bibfield{author}{\bibinfo{person}{Keiron O'Shea} {and} \bibinfo{person}{Ryan
  Nash}.} \bibinfo{year}{2015}\natexlab{}.
\newblock \showarticletitle{An introduction to convolutional neural networks}.
\newblock \bibinfo{journal}{\emph{arXiv preprint arXiv:1511.08458}}
  (\bibinfo{year}{2015}).
\newblock


\bibitem[Pazzani(1999)]%
        {pazzani1999framework}
\bibfield{author}{\bibinfo{person}{Michael~J Pazzani}.}
  \bibinfo{year}{1999}\natexlab{}.
\newblock \showarticletitle{A framework for collaborative, content-based and
  demographic filtering}.
\newblock \bibinfo{journal}{\emph{Artificial intelligence review}}
  \bibinfo{volume}{13} (\bibinfo{year}{1999}), \bibinfo{pages}{393--408}.
\newblock


\bibitem[Raffel et~al\mbox{.}(2020)]%
        {raffel2020exploring}
\bibfield{author}{\bibinfo{person}{Colin Raffel}, \bibinfo{person}{Noam
  Shazeer}, \bibinfo{person}{Adam Roberts}, \bibinfo{person}{Katherine Lee},
  \bibinfo{person}{Sharan Narang}, \bibinfo{person}{Michael Matena},
  \bibinfo{person}{Yanqi Zhou}, \bibinfo{person}{Wei Li}, {and}
  \bibinfo{person}{Peter~J Liu}.} \bibinfo{year}{2020}\natexlab{}.
\newblock \showarticletitle{Exploring the limits of transfer learning with a
  unified text-to-text transformer}.
\newblock \bibinfo{journal}{\emph{The Journal of Machine Learning Research}}
  \bibinfo{volume}{21}, \bibinfo{number}{1} (\bibinfo{year}{2020}),
  \bibinfo{pages}{5485--5551}.
\newblock


\bibitem[Ramos et~al\mbox{.}(2003)]%
        {ramos2003using}
\bibfield{author}{\bibinfo{person}{Juan Ramos} {et~al\mbox{.}}}
  \bibinfo{year}{2003}\natexlab{}.
\newblock \showarticletitle{Using tf-idf to determine word relevance in
  document queries}. In \bibinfo{booktitle}{\emph{Proceedings of the first
  instructional conference on machine learning}}, Vol.~\bibinfo{volume}{242}.
  Citeseer, \bibinfo{pages}{29--48}.
\newblock


\bibitem[Schafer et~al\mbox{.}(2007)]%
        {schafer2007collaborative}
\bibfield{author}{\bibinfo{person}{J~Ben Schafer}, \bibinfo{person}{Dan
  Frankowski}, \bibinfo{person}{Jon Herlocker}, {and} \bibinfo{person}{Shilad
  Sen}.} \bibinfo{year}{2007}\natexlab{}.
\newblock \showarticletitle{Collaborative filtering recommender systems}.
\newblock \bibinfo{journal}{\emph{The adaptive web: methods and strategies of
  web personalization}} (\bibinfo{year}{2007}), \bibinfo{pages}{291--324}.
\newblock


\bibitem[Sherstinsky(2020)]%
        {sherstinsky2020fundamentals}
\bibfield{author}{\bibinfo{person}{Alex Sherstinsky}.}
  \bibinfo{year}{2020}\natexlab{}.
\newblock \showarticletitle{Fundamentals of recurrent neural network (RNN) and
  long short-term memory (LSTM) network}.
\newblock \bibinfo{journal}{\emph{Physica D: Nonlinear Phenomena}}
  \bibinfo{volume}{404} (\bibinfo{year}{2020}), \bibinfo{pages}{132306}.
\newblock


\bibitem[Son and Kim(2017)]%
        {son2017content}
\bibfield{author}{\bibinfo{person}{Jieun Son} {and} \bibinfo{person}{Seoung~Bum
  Kim}.} \bibinfo{year}{2017}\natexlab{}.
\newblock \showarticletitle{Content-based filtering for recommendation systems
  using multiattribute networks}.
\newblock \bibinfo{journal}{\emph{Expert Systems with Applications}}
  \bibinfo{volume}{89} (\bibinfo{year}{2017}), \bibinfo{pages}{404--412}.
\newblock


\bibitem[Touvron et~al\mbox{.}(2023)]%
        {touvron2023llama}
\bibfield{author}{\bibinfo{person}{Hugo Touvron}, \bibinfo{person}{Thibaut
  Lavril}, \bibinfo{person}{Gautier Izacard}, \bibinfo{person}{Xavier
  Martinet}, \bibinfo{person}{Marie-Anne Lachaux},
  \bibinfo{person}{Timoth{\'e}e Lacroix}, \bibinfo{person}{Baptiste
  Rozi{\`e}re}, \bibinfo{person}{Naman Goyal}, \bibinfo{person}{Eric Hambro},
  \bibinfo{person}{Faisal Azhar}, {et~al\mbox{.}}}
  \bibinfo{year}{2023}\natexlab{}.
\newblock \showarticletitle{Llama: Open and efficient foundation language
  models}.
\newblock \bibinfo{journal}{\emph{arXiv preprint arXiv:2302.13971}}
  (\bibinfo{year}{2023}).
\newblock


\bibitem[Van~Meteren and Van~Someren(2000)]%
        {van2000using}
\bibfield{author}{\bibinfo{person}{Robin Van~Meteren} {and}
  \bibinfo{person}{Maarten Van~Someren}.} \bibinfo{year}{2000}\natexlab{}.
\newblock \showarticletitle{Using content-based filtering for recommendation}.
  In \bibinfo{booktitle}{\emph{Proceedings of the machine learning in the new
  information age: MLnet/ECML2000 workshop}}, Vol.~\bibinfo{volume}{30}.
  Barcelona, \bibinfo{pages}{47--56}.
\newblock


\end{thebibliography}

\end{document}